\documentclass[a4paper]{jpconf}
\usepackage{graphicx}
\usepackage{wrapfig}
\usepackage{amsmath}
\usepackage{hyperref}

\usepackage{lineno}
\begin{document}
\title{Machine Learning approach to boosting neutral particles identification in the LHCb calorimeter}
\author{A Boldyrev$^{1}$, V Chekalina$^{1}$, F Ratnikov$^{1,2}$ on behalf of the LHCb collaboration}
\address{$^1$ National Research University Higher School of Economics, Laboratory of Methods for Big Data Analysis, 3 Kochnovsky Proezd, Moscow 125319, Russia}
\address{$^2$ The Yandex School of Data Analysis, 11/2 Timura Frunze St., Moscow 119021, Russia}

\ead{alexey.boldyrev@cern.ch}

\begin{abstract}
We present a new approach to identification of boosted neutral particles using Electromagnetic Calorimeter (ECAL) of the LHCb detector. The identification of photons and neutral pions is currently based on the geometric parameters which characterise the expected shape of energy deposition in the calorimeter. This allows to distinguish single photons in the electromagnetic calorimeter from overlapping photons produced from high momentum $\pi^0$ decays. The novel approach proposed here is based on applying machine learning techniques to primary calorimeter information, that are energies collected in individual cells around the energy cluster. This method allows to improve separation performance of photons and neutral pions and has no significant energy dependence.
\end{abstract}

\section{Introduction}
A few important analyses at LHCb need to reconstruct energetic photons in the final state. The notable example is the analyses of radiative B decays which serve as a sensitive probe for various extensions of the Standard Model. Such decays are affected by the backgrounds from B decays where a photon is substituted by a neutral pion ($\pi^0$). Since the $\pi^0$ can be produced with high momentum in the laboratory frame, the two photons from its decay can produce overlapping clusters in the calorimeter, and can thus be misinterpreted as the single energetic photon. It is thus very important to be able to distinguish between the high-momentum photons and $\pi^0$ using the shape of the calorimeter clusters. This paper describes the novel approach to identification of boosted neutral particles, evaluation of its performance obtained on MC samples, and discusses specific issues when transferring discriminative models from simulation to real world.
% The  machine learning model employs gradient boosting trees approach which is widely used nowadays, and separates $\pi^0$ and photon responses from ``first principles''. This approach allowed to significantly improve separation performance score on simulated data, reducing primary photons fake rate by factor of four. In this paper we describe the approach, evaluate its performance obtained on MC samples, and discuss specific issues when transferring discriminative models from simulation to real world.

\section{Electromagnetic Calorimeter}
The LHCb calorimeter system performs several tasks, providing the first level trigger with high transverse momentum photon, electron and hadron candidates, measuring their energies and positions and performing the separation between photons, electrons and hadrons~\cite{LHCbMain}. The LHCb ECAL~\cite{LHCbMain2} is based on the Shashlik technology of alternating scintillating tiles and lead plates, preceded by a Preshower (PS) and contains 1536/1792/2688 cells in its inner/middle/outer regions, respectively.

When the ECAL cell has an excess in energy deposition (compared to the adjacent cells) it will originate a cluster according to the following procedure~\cite{main}. Energy deposits in ECAL cells are clusterised applying a 3$\times$3 cell pattern around the local maximum of energy deposition or seed cell. Consequently, the seed cells of the reconstructed clusters are always separated by at least one cell. The transverse energy of the seed cell is required to be larger than 50 MeV. Neutral clusters are identified as those clusters that do not match to charged tracks extrapolated to the calorimeter surface. For each track-cluster pair, $\chi^2_{2D}$ is obtained taking into account: the position of the point of intersection of the extrapolated track from the calorimeter, the covariance matrix of track parameters, the position of the barycenter of the cluster and the matrix of the second moments of the cluster.

The signature of a $\pi^0$ decay in the ECAL depends on the kinematics of the two photons. Low-momentum $\pi^0$ produces two separated clusters. Such $\pi^0$ are classified as resolved $\pi^0$ and its reconstruction is based on the invariant mass of the photon pair. High-momentum $\pi^0$ produces a single cluster and it is classified as a merged $\pi^0$. The transition region between high- and low-momentum $\pi^0$ occurs around 2~GeV/c.

The discriminating features separating photon and merged $\pi^0$ clusters are based on the cluster shape. Merged $\pi^0$ clusters are expected to be elongated and asymmetric due to residual offset between two photons, while genuine photon clusters are expected to be more symmetrical. Full description of the variables used for the discrimination between photons and merged $\pi^0$, which we call shape-based approach (or \textit{baseline}), can be found in~\cite{main}.

% Around 88\% of the photons in ECAL originate from a $\pi^0$ decay. The signature in the ECAL depends on its kinematics, the higher the momentum of the $\pi^0$ is the closer the two photons are at the entry of the calorimeter. These two photons can then either produce two separated clusters or share a single cluster in which their individual signals are not clearly distinguishable.

%Cluster shape variables combined with additional information from the PS detector and a 2-layer perceptron (MLP) as classifier are used for the $\gamma/\pi^0$ separation.

% Since the pion background is large for the radiative decays the use of an efficient PID variable is crucial.

\section{Current Machine Learning approaches to Particle Identification}
Particle identification algorithms can be based on multivariate classifiers which allow a straightforward combination of information originating from different subdetectors (i.e. PS and ECAL) into a discriminant output. Specifically to our problem, it allows extending the baseline method to consider track-cluster matching or the shape of the neutral cluster.

\begin{figure} [hbt]
\centering
    \begin{minipage}{.5\textwidth}
        \centering
        \includegraphics[width=\linewidth]{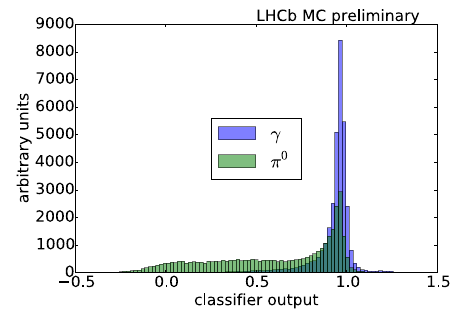}
        \caption{\label{fig:classifier_1} Baseline output (IsPhoton variable)}
    \end{minipage}%
\centering
    \begin{minipage}{.5\textwidth}
        \centering
        \includegraphics[width=\linewidth]{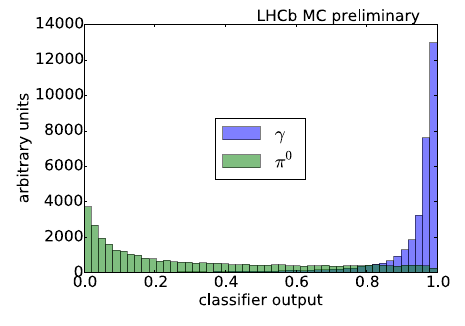}
        \caption{\label{fig:classifier_2} XGBoost approach output}
    \end{minipage}%
\end{figure}

Particle identification algorithms are trained to separate photons signatures from hadrons and electrons, which may have passed the neutral cluster selection, and high-energy $\pi^0$. The classifier output of both baseline and XGBoost approaches is displayed on Figure~\ref{fig:classifier_1} and~\ref{fig:classifier_2} using simulated data samples for both training and performance evaluation. The performance of both approaches is presented as the dashed line in Figure~\ref{fig:roc}.

Due to MC/data disagreement in the input variables used to build the $\gamma/\pi^0$ separation variable,
discrepancies are also expected in the output of the method. To get a true estimate of the selection efficiency for a given cut on the classifier output, calibration samples from real data are needed.

\section{Calibration samples}

In order to calibrate the performance of the discriminative variable,
$B^{0} \rightarrow K \pi \gamma$ reconstructed events
are used as calibration samples for photons and $B^{0} \rightarrow K \pi \pi^0$ for $\pi^0$. The kinematically similar decay is chosen to prevent a biased output with respect to the particle energy.
To prove stability of the method, additional $\pi^0$ samples from $B^0 \rightarrow J/\psi K^{\ast}$ (where $K^{\ast} \rightarrow K \pi^0$) are used. Due to the high muon trigger efficiency, the presence of $J/\psi$ mesons decaying into pair of muons provides high detection efficiency of the selected decays of $B^0$ mesons. We required $K$ and $\pi$ to have transverse momentum $p_T$ > 500 GeV/c and the vertex quality of the $K$ and $\pi$ tracks forming the $K^{\ast}$ candidate to be $\chi^2(K^{\ast})$ < 9. %Same selection criteria is applied for each samples.
We considered energy clusters with transverse energy $E_T > 2$~GeV.

\section{New approach}
%taken from Chekalina_2018_J._Phys._Conf._Ser._1085_042036
To separate photons and merged $\pi^0$, we take into account energies in a 5$\times$5 ECAL window and the PS cells around the cell seed. However we do not build any sophisticated features based on physics considerations, but rather consider plain values of energy allocated in every cell of the 5$\times$5 matrix in both ECAL and PS as features.
These are 50 plain features to be used by the classifier. The technical aspects of the proposed approach was discussed in detail in the conference paper~\cite{Chekalina_2018}. In this study, the inner region of the ECAL is used as a reference for quantitative comparison of the classifiers.

\begin{figure} [hbt]
\centering{
\includegraphics[width=0.75\linewidth]{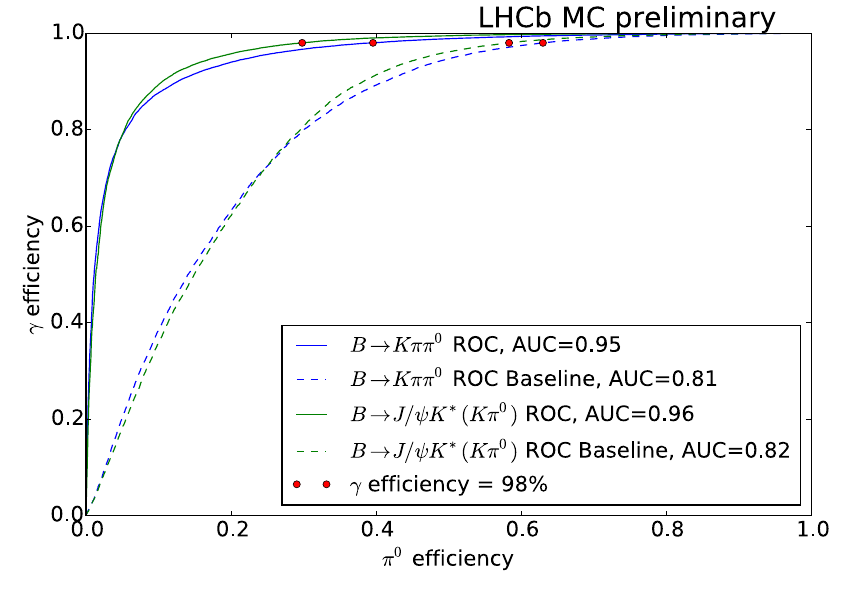}
\caption{\label{fig:roc} Receiver Operating Characteristic (ROC) curves for the baseline (dashed line) and new approach (solid line). Different colours refer to different test samples}
}
\end{figure}

% We selected the XGBoost tool~\cite{xgboost} with 6000 trees with depth 3. Figure~\ref{fig:roc} demonstrates performance of the new approach in
% comparison with the baseline one. The score under Receiver operating characteristic curve improves from 0.89
% for the baseline up to 0.97 in new approach. Considering 98\% photon
% efficiency, new approach reduces fake rate from about
% 60\% to about 30\%.

% Flatness is an important characteristics of the classifier, as it
% directly affects systematic uncertainties for physics analyses.
% Figure~\ref{fig:roc_et} demonstrates that new approach also have very
% good flatness in $E_T$.

\begin{figure} [hbt]
\centering{
\includegraphics[width=\linewidth]{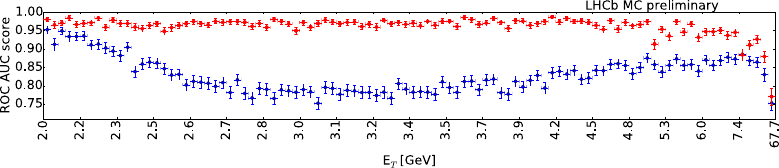}
\caption{\label{fig:roc_et} Baseline (blue) and BDT approach (red) model quality as a function of transverse energy}
}
\end{figure}

\section{Classifier selection and performance}
We used several classifiers based on Neural Network and Boosted Decision Tree architectures.

For possible NN architecture, we build two parallel branches fed from preprocessed information from the ECAL and PS. We found that using PS branch with smaller number of units together with ECAL branch could improve quality of the classifier. Among the hyperparameters for the NN are: number of layers and number of units in each layer, method of regularisation, activation function. To achieve best quality of the classifier, we tuned all the hyperparameters leaving the network topology unchanged. To determine the number of training epochs, we monitored the train-test score curve and observed when it reached the plateau. As a result, we used 3-fold cross-validation and trained the network up to 5000 epochs. The insufficiently complicated structure of input features leads to a degradation of classifier quality with NN configurations with 3 and 4 hidden layers. % The training algorithm and network structure (number of nodes) is not specified.

For the BDT approach, we assayed XGBoost, CatBoost and LightGBM classifiers.
For each classifier, we find the best configuration by tuning the boosting parameters using ModelGym~\cite{modelgym}. The different classifiers qualities are found to be very close to each other.

As BDT-based classifiers demonstrate better performance for the problem than the best NN configuration, we choose the BDT approach.
The default XGBoost~\cite{xgboost} was selected (estimators = 2000, learning rate = 0.05, max depth = 5, min child weight = 2).
We used XGBoost with 6000 trees with depth set to 3. Figure~\ref{fig:roc} demonstrates the performance of the new approach in comparison to the baseline one.
The score in the ROC curve improves from 0.89
for the baseline to up to 0.97 in the new approach. Considering 98\% photon efficiency, the new approach reduces fake rate from about 60\% to about 30\%.

An unbiased behaviour with respect to energy is an important characteristics of the classifier, as it
directly affects systematic uncertainties for physics analyses.
As one can see in Figure~\ref{fig:roc_et}, the new approach displays a flat efficiency profile with respect to $E_T$.

\section{Conclusion}
We developed a new procedure to separate photons from merged $\pi^0$. The new approach shows good performance on simulated data and reasonably good performance using only the ECAL energy deposits. The proposed classifier shows negligible energy dependency which can be useful to estimate systematic uncertainties for physics analyses.

%The advantage of the approach is that it can show stable performance under different running conditions.
However, careful consideration of the simulated data used in efficiency evaluations should be taken. Since the new approach uses solely energy deposits in the ECAL, it can be considered as a necessary part of future neutral particle identification tools.

\section{Acknowledgements}
The research leading to these results has received funding from Russian Science Foundation under grant agreement n$^{\circ}$~19-71-30020.

\nocite{*}

% \bibliography{apssamp}% Produces the bibliography via BibTeX.

% \section{References}
% \bibliography{refs}{}

\section*{References}
\begin{thebibliography}{9}
\bibitem{LHCbMain}
LHCb collaboration, A. A. Alves Jr. {\it et al.}, {\it The LHCb detector at the
LHC}, JINST {\bf 3} (2008) S08005
\bibitem{LHCbMain2}
LHCb collaboration, R. Aaij {\it et al.}, {\it LHCb detector performance}, Int. J. Mod. Phys.
{\bf A30} (2015) 1530022, arXiv:1412.6352
\bibitem{main}
 M. Calvo, E. Cogneras, O. Deschamps, M. Hoballah, {\it  A tool for $\gamma/\pi^0$ separation at high energies}, LHCb-PUB-2015-016
% \bibitem{Chekalina_2018}
% Viktoria Chekalina and Fedor Ratnikov 2018 {\it J. Phys.: Conf. Ser.} {\bf 1085} 042036
\bibitem{Chekalina_2018}
V. Chekalina and F. Ratnikov, {\it Machine Learning approach to $\gamma/\pi^0$ separation in the LHCb calorimeter}, Phys.: Conf. Ser. {\bf 1085} (2018) 042036
% \bibitem{tesla}
% LHCb Collaboration, R. Aaij {\it et al.}, {\it Tesla: an application for real-time data analysis in High Energy Physics}, Comput. Phys. Commun. {\bf 208} (2016) 35-42
\bibitem{modelgym}
Model Gym Project 2017 Model Gym [software] Available from \url{https://github.com/yandexdataschool/modelgym}
[accessed 2019-05-14]
\bibitem{xgboost}
XGBoost: A Scalable Tree Boosting System. {\tt arXiv:1603.02754
  [cs.LG]},  March 2016
% \bibitem{xgboost-sw}
% XGBoost Project 2018 XGBoost [software] Available from \url{https://github.com/dmlc/xgboost}
% [accessed 2019-05-20]
% \bibitem{splot}
% M. Pivk and F. R. Le Diberder, {\it SPlot: A Statistical tool to unfold data distributions}, Nucl. Instrum.
% Meth. {\bf A 555} (2005) 356
\end{thebibliography}
% \bibliographystyle{iopart-num}

\end{document}